\documentclass{article}
\usepackage[utf8]{inputenc}
\usepackage{graphicx}
\usepackage{jheppub}
\usepackage{color}
\usepackage{amsfonts}
\usepackage{latexsym}
\usepackage{amsmath,amssymb}
\usepackage{verbatim}
\usepackage{comment}
\usepackage{hyperref}
\numberwithin{equation}{section}
\usepackage{mathtools}
\usepackage{tcolorbox}

 \newcommand{\be}{\begin{equation}}
\newcommand{\ee}{\end{equation}}
\newcommand{\bea}{\begin{eqnarray}}
\newcommand{\eea}{\end{eqnarray}}

\newcommand{\ba}{\begin{aligned}}
\newcommand{\ea}{\end{aligned}}
\newcommand{\p}{\partial}

\title{Static electromagnetic Love tensors of 5-dimensional Myers-Perry black holes}

\author[a,b]{Boyang Yu}

\affiliation[a]{School of Mathematics and Maxwell Institute for Mathematical Sciences, University of Edinburgh, Edinburgh EH9 3FD, UK}
\affiliation[b]{Center for High Energy Physics, Peking University, No.5 Yiheyuan Rd, Beijing 100871, P.R. China}

\emailAdd{v1byu33@ed.ac.uk}

\abstract{We study the separable master equations for the electromagnetic and  gravitational perturbations in five-dimensional Myers-Perry black holes. In the static limit, while the master equation for the electric polarization of the Maxwell field reduces to that of a massless scalar field, the magnetic polarization and gravitational perturbation yield Heun equations for both its angular and radial components. Remarkably, these Heun equations fall into a special class that admits exact analytic solutions in terms of hypergeometric functions. We reconstruct the gauge field using master fields and study its asymptotic behavior. When expanding the result in the basis of modified spherical harmonics, we find modes with higher angular momentum arise in response to the excitation of sources with lower angular momentum. The static tidal Love tensor that characterizes such mixing structure of the response can be computed iteratively. We also discuss the possible near zone approximation of the master equations for the magnetic polarization.

}

\begin{document}

\maketitle
\flushbottom

\section{Introduction}

Understanding wave propagation in black hole backgrounds is essential for exploring a wide range of theoretical and phenomenological phenomena in gravitational physics. The study of wave equations in curved spacetimes, such as those describing scalar, electromagnetic, and gravitational perturbations, enables the computation of key physical quantities like quasinormal modes (QNMs), greybody factors, and holographic thermal correlators.
Another important motivation for studying black hole perturbations is the extraction of static response coefficients, notably the tidal Love numbers, initially introduced by Love in the context of Newtonian gravity in 1909 \cite{Love:1909}. These dimensionless parameters measure the susceptibility of compact spherically symmetric objects, such as neutron stars and black holes, to tidal deformation induced by external gravitational fields \cite{Binnington:2009bb,Damour:2009vw}. Generalizations to spinning objects such as Kerr black holes have also been investigated \cite{Goldberger:2020fot,Charalambous:2021kcz,LeTiec:2020bos}.

Recent developments in effective field theory (EFT) have emphasized the interpretation of black holes as point particles characterized by multipole moments from the perspective of a distant observer \cite{Goldberger:2004jt,Goldberger:2005cd,Porto:2016pyg,Porto:2016zng}. Within this EFT framework, Love numbers correspond precisely to finite-size corrections encoded in the Wilsonian effective action, providing insights into black hole microstructure and symmetries \cite{Porto:2016pyg,Charalambous:2021kcz}. To compute the Love numbers, one needs to solve related wave equations with proper boundary conditions imposed in the interior and expand the solution in the asymptotic region. 

Under suitable approximations, wave equations in black hole backgrounds could exhibit enhanced symmetries known as hidden symmetries.  A well-known example is the near-extremal Kerr black hole where the Teukolsky equation features local $SL(2,\mathbb R)_L\times SL(2,\mathbb R)_R$ symmetry, leading to the Kerr/CFT proposal \cite{Castro:2010fd}. Nevertheless, this hidden symmetry is not globally well defined. \cite{Charalambous:2021kcz} considered a slightly different approximation of the Teukolsky equation in the general Kerr background and found a hidden $SL(2,\mathbb R)$ symmetry, which they call the Love symmetry. By solving the Teukolsky equation, 
it was found that the Love numbers would vanish under certain conditions, which can be understood using the representation of the Love symmetry algebra. Love symmetries and their usage in studying Love numbers for massless scalar fields were also discussed in the five-dimensional Myers-Perry black hole \cite{Charalambous:2023jgq} and black p-branes \cite{Charalambous:2025ekl}.

In this work, we study the  electromagnetic and gravitational perturbations in the five-dimensional Myers-Perry black hole. Separability of the wave equations is crucial for our analysis. The separability of Maxwell's equations in the five-dimensional Myers-Perry geometry has already been shown in \cite{Lunin:2017drx}, and  the wave equation for the vector polarization of gravitational wave was also shown to be separable recently. Technically, the author showed that the electromagnetic perturbation and the vector type of the gravitational perturbations can be reconstructed from scalar master fields. The Maxwell equations and linearized Einstein equations then become  wave equations for corresponding master fields, which are proved to be separable.
These two types of separable excitations will be the focus of this work. We will focus on the static limit of these separable wave equations which turn out to be Heun equations. 

The Heun equation and its confluent forms arise commonly in wave equations associated with black hole backgrounds \cite{Hortacsu:2011rr}. They can be identified with the semiclassical BPZ equations that govern five-point functions featuring one degenerate operator insertion in Liouville CFT \cite{Belavin:1984vu}. Consequently, the connection problem for  Heun type equations can be treated as a connection problem for semiclassical conformal blocks. Thanks to the  Alday–Gaiotto–Tachikawa(AGT) correspondence \cite{Alday:2009aq}, the relevant semiclassical conformal blocks
 can be expressed in terms of the partition function of the four-dimensional
gauge theory in the Nekrasov-Shatashvili limit \cite{Nekrasov:2003rj,Nekrasov:2002qd,Nekrasov:2009rc}. Based on these works, the connection formulae for the Heun equation and its confluences have been obtained \cite{Bonelli:2022ten,Lisovyy:2022flm}. These results have wide applications and allow one to compute important quantities related to black hole scattering problems, such as greybody factors, quasinormal modes, holographic thermal correlators and Love numbers \cite{Bonelli:2021uvf,Pereniguez:2021xcj,Dodelson:2022yvn,Consoli:2022eey,Dodelson:2023vrw,Bucciotti:2023ets}.   Nevertheless, results for these physical quantities are usually very complicated and studied numerically. However, there are special cases when the solution to the Heun equation can be written as a finite sum of hypergeometric functions. This occurs when the parameters of the Heun equation satisfy certain constraints. This seems to be a very special case and purely mathematical. However, when studying the static Love number of the scalar type of gravitational perturbation in higher dimensional Schwarzschild black hole, it is found that the perturbation satisfies the Zerilli equation which can be transformed into a Heun equation after some field redefinitions \cite{Hui:2020xxx}. 
Remarkably, one of the singular points of the resulting Heun equation is removable
and the equation can be solved analytically in terms of hypergeometric functions.
The same phenomenon has also been observed when studying 
perturbations around four dimensional Reissner–Nordström black holes \cite{Rai:2024lho}.
For the static perturbations we study in this paper, we find that the corresponding wave equations share the same feature: they are Heun equations that can be solved analytically. With these analytic solutions at hand, we reconstruct the perturbations and analyze the tidal response within the framework of effective theory (EFT). 

The paper is organized as follows. Section 2 reviews the five-dimensional Myers-Perry black hole geometry and the separability of wave equations for the gauge field and the vector polarization of the gravitational perturbation. In section 3, we solve the static limit of the wave equations. Then we study the asymptotic behavior of these static solutions and compute the running Love numbers.
In section 4, we reconstruct electromagnetic perturbations using the master fields we have solved and identify the tensor structure of tidal responses. We also explicitly compute the values of Love tensors for the first few orders. In section 5, we move away from the static limit and comment on the possible near zone approximations for the wave equations. We present a short summary of our results and provide some open questions that can be studied in the future. A brief review of Heun equation and modified spherical harmonics is presented in the appendices.

\section{Review of Myers-Perry black holes and wave equations}
In this section, we briefly review the geometry of 5d Myers-Perry black hole and the separable ansatz for both electromagnetic and gravitational waves following \cite{Lunin:2017drx,Lunin:2025yth}. 
\subsection{ Myers–Perry black hole in 5d}
The metric of the Myers–Perry black hole in $d=2n+1$ dimension is presented as \cite{Myers:1986un}
\be
ds^2=-dt^2+\frac{Mr^2}{FR}\left(dt+\sum_{i=1}^na_i\mu_i^2d\phi_i\right)^2+\frac{FRdr^2}{R-Mr^2}+\sum_{i=1}^n(r^2+a_i^2)(d\mu_i^2+\mu_i^2d\phi_i^2)\,,
\ee
where
\be
R=\prod_{k=1}^n(r^2+a_k^2)\,,\quad FR=r^2\prod_{k=1}^n(r^2+x_k^2)\,,\quad \mu_i^2=\frac{1}{c_i^2}\prod_{k=1}^{n-1}(a_i^2-x_k^2)
\ee
with $c_i^2=\prod_{k\neq i}(a_i^2-a_k^2)$. For $n=2$, setting $a_1=-a,a_2=-b$, we have $c_1^2=-c_2^2=a^2-b^2$. Thus, coordinates $\mu_1$ and $\mu_2$ are given by
\be
\mu_1^2=\frac{a^2-x_1^2}{a^2-b^2}\,,\quad \mu_2^2=\frac{b^2-x_1^2}{b^2-a^2}\,,
\ee
which satisfy $\mu_1^2+\mu_2^2=1$.
We choose the parametrization $\mu_1=\sin\theta,\mu_2=\cos\theta$ so that $x_1$ becomes
\be
x_1=\sqrt{(a\cos\theta)^2+(b\sin\theta)^2}\,.
\ee
Finally, letting $\phi_1=\phi,\phi_2=\psi$, the explicit form of the metric in coordinates $(t,r,\theta,\phi,\psi)$ is given by 
\be\ba\label{MPmetric}
ds^2=&-dt^2+\frac{M}{\Sigma}\left(dt-a\sin^2\theta d\phi-b\cos^2\theta d\psi\right)^2+\frac{r^2\Sigma}{\Delta}dr^2+\Sigma d\theta^2
\\&+(r^2+a^2)\sin^2\theta d\phi^2+(r^2+b^2)\cos^2\theta d\psi^2\,,
\ea\ee
with $\Delta=R-Mr^2,\Sigma=r^2+x_1^2$. The metric \eqref{MPmetric} describes a five-dimensional rotating black hole with mass $M$ and two angular velocities $a,b$.
It is convenient to introduce local frames, which play an important role in studying the separability of electromagnetic wave equations, and they are given by
\be\ba
l_\pm&=\pm\frac{R}{r}\p_t+\frac{\Delta}{r}\p_r\pm\frac{aR_b}{r}\p_\phi\pm\frac{bR_a}{r}\p_\psi\,,
\\ m_\pm&=\pm\frac{i(a^2-b^2)\cos\theta\sin\theta}{x_1}\p_t+\p_\theta\pm\frac{ia\cot\theta}{x_1}\p_\phi\mp\frac{ib\tan\theta}{x_1}\p_\psi\,,
\\n&=ab\p_t+b\p_\phi+a\p_\psi\,,
\ea
\ee
with $R_a=r^2+a^2,R_b=r^2+b^2$. The inverse metric can be written in terms of above frames as
\be
g^{\mu\nu}\p_\mu\p_\nu=\frac{1}{\Sigma\Delta}l_+^\mu l_-^\nu\p_\mu\p_\nu+\frac{1}{\Sigma}m_+^\mu m_-^\nu\p_\mu\p_\nu+\frac{1}{r^2x_1^2}n^\mu n^\nu\p_\mu\p_\nu\,.
\ee

\subsection{Separable wave equations}
\paragraph{Scalar field}
The massless scalar field $\Psi^{S}$ satisfies the Klein-Gordon equation
\be\label{eq:KG}
\Box\Psi^{\text{S}}=0.
\ee
If we separate $\Psi^{\text{S}}$ as
\be\label{Psi}
\Psi^{S}=e^{i\omega t+im\phi+in\psi}\Phi^{S}_{\omega\ell mn}(r)S^{S}_{\omega\ell mn}(\theta)
\ee 
with $\ell$ being the orbital number which characterizes the separation constant,
then \eqref{eq:KG} can be written into two decoupled second order ODEs for $\Phi^S_{\omega\ell mn}(r)$ and $S^S_{\omega\ell mn}(\theta)$, with each defining an eigenvalue problem for the separation constant. The explicit form of such decoupled ODEs will be summarized later.
\paragraph{Electromagnetic field}

The separability of 
Maxwell's equations for generic  electromagnetic perturbations 
     $A_\mu$ in $d>4$ dimensional spacetime requires the background spacetime to be Kundt, which is defined to be algebraically special and admit a null geodesic congruence with vanishing expansion, shear, and twist \cite{Durkee:2010qu}. Unfortunately, Myers-Perry black hole is not Kundt. This indicates that separating the gauge field into the form \eqref{Psi} is inconsistent with the Maxwell's equations.
 Nevertheless, a remarkable work by \cite{Lunin:2017drx} shows that the Maxwell's equations can be transformed into a separable master equation for a scalar field which relates to $A_\mu$ via the following ansatz
\be\label{ansatz-general}
l^\mu_\pm A_\mu=G_\pm(r)l^\mu_\pm\p_\mu\Psi^{\text{EM}}\,,\quad m^\mu_\pm A_\mu=F_\pm(\theta)m^\mu_\pm\p_\mu\Psi^{\text{EM}}\,,\quad n^\mu A_\mu=\lambda \Psi^{\text{EM}}\,,
\ee
where $G_\pm,F_\pm$  and $\lambda $ are determined by the Maxwell's equations.  
When the master field $\Psi^{\text{EM}}$ is separated as \eqref{Psi},  Maxwell's equations result in two decoupled ODEs for the radial and angular functions.  There are two types of choices for $G_\pm,F_\pm$ and $\ell $ with $\Psi^{\text{EM}}$ being called electric polarization and magnetic polarization correspondingly. For the electric polarization, the factors are chosen to be
\be\label{ansatz-e}
G_\pm(r)=\pm\frac{r}{1\pm i\mu r}\,,\quad F_\pm(\theta)=\mp\frac{ix_1}{1\pm\mu x_1}\,,\quad\lambda =0\,,
\ee
while for the magnetic polarization, the factors are chosen to be
\be
G_\pm(r)=\pm\frac{1}{r\pm i\mu}\,,\quad F_\pm(\theta)=\pm\frac{i}{x_1\mp\mu}\,,\quad\lambda=\frac{ab\omega+an+bm}{\mu}\,.
\ee
It turns out that the decoupled angular and radial equations for both scalar  and electromagnetic fields can be presented uniformly as follows
\be\ba\label{eq}
{}&\frac{D^\alpha_\theta}{\sin2\theta}\frac{d}{d\theta}\left[\frac{\sin2\theta}{D_\theta^\alpha}\frac{d}{d\theta}S^\alpha_{\omega\ell mn}\right]+\left[\frac{2\Lambda^\alpha}{D_\theta^\alpha}+\omega^2x_1^2-\frac{m^2}{\sin^2\theta}-\frac{n^2}{\cos^2\theta}+C^\alpha\right]S^\alpha_{\omega \ell mn}=0\,,
\\{}&\frac{D^\alpha_r}{r}\frac{d}{dr}\left[\frac{\Delta}{rD^\alpha_r}\frac{d}{dr}\Phi^\alpha_{\omega\ell mn}\right]+\left[-\frac{2\Lambda^\alpha}{D^\alpha_r}+(\omega r)^2+\frac{m^2d_a}{R_a}+\frac{n^2d_b}{R_b}-C^\alpha+\frac{MRW^2}{\Delta}\right]\Phi^\alpha_{\omega \ell mn}=0\,,
\ea
\ee
where $\alpha\in\{S,E,M\}$ labels different kinds of waves and
\be
 W=\omega+\frac{am}{R_a}+\frac{bn}{R_b}\,,\quad d_a=-d_b=a^2-b^2\,.
\ee
Values of $D_\theta,D_r$ and $\Lambda,C$ for different kinds of perturbations are summarized below
\be\ba\label{choice}
\text{scalar:}&\quad D^S_r=D^S_\theta=1\,;\\
\text{electric:}&\quad D^E_r=1+(\mu r)^2\,,\quad D^E_\theta=1-(\mu x_1)^2\,,\quad C^E=(\mu ab\tilde{\Omega})^2+\tilde C\,,\\
&\Lambda^E=\omega\mu^3\left(\frac{1}{\mu^2}-\frac{am}{\omega}-a^2\right)\left(\frac{1}{\mu^2}-\frac{bn}{\omega}-b^2\right)-\frac{\mu^3abmn}{\omega}\,;\\
\text{magnetic:}&\quad D^M_r=-1-\frac{r^2}{\mu^2}\,,\quad D^M_\theta=-1+\frac{x_1^2}{\mu^2}\,,\quad C^M=\frac{(ab\tilde{\Omega})^2}{\mu^2}+\tilde C\,,\\
&\Lambda^M=\frac{\omega}{\mu^3}\left({\mu^2}-\frac{am}{\omega}-a^2\right)\left(\mu^2-\frac{bn}{\omega}-b^2\right)-\frac{abmn}{\mu^3\omega}\,,
\ea
\ee
with
\be
\tilde C=-\omega^2(a^2+b^2)-2\omega(am+bn)\,,\quad \tilde\Omega=\omega+\frac{m}{a}+\frac{n}{b}\,.
\ee
Consequently, the wave equations \eqref{eq} define  eigenvalue problems for $C_0\equiv C^S+2\Lambda^S$ for the scalar field, while for electric and magnetic perturbations, \eqref{eq} define eigenvalue problems for the parameter $\mu$. 
\paragraph{Gravitational field of vector type}
In the recent work \cite{Lunin:2025yth}, the author studied gravitational wave equations in the Myers–Perry geometry of arbitrary dimensions, with at least one rotation parameter set to zero. The gravitational wave in such geometry can be classified into three types: scalar, vector, and tensor types \cite{Kodama:2007ph,Ishibashi:2011ws}. In particular,
for the five-dimensional Myers-Perry black hole with one rotating parameter set to 0, \cite{Lunin:2025yth} showed that the wave equation for the vector mode is also separable. We set $b=0$ in the following.
Similar to the case of the electromagnetic field, the metric perturbations for the vector mode are related to a scalar master field $\Psi^V$ via the following ansatz
\be\label{ansatz-grav}
l_\pm^\mu n^\nu h_{\mu\nu}=ar^2x_1^2\left(\frac{r}{r\pm i\mu}+\zeta\right)l_\pm^\mu\p_\mu\Psi^V\,,\quad m_\pm^\mu n^\nu h_{\mu\nu}=ar^2x_1^2\left(\frac{x_1}{x_1\mp\mu}+\zeta\right)m_\pm^\mu\p_\mu\Psi^V\,
\ee
with the remaining projections being zero. Moreover, $\Psi^V$ is independent of the coordinate $\psi$.
Parameter $\mu$ is the separation constant and $\zeta$ describes the pure gauge with the choice $\zeta=-1$ corresponding to the  Lorenz gauge \cite{Frolov:2018pys,Krtous:2018bvk}. Using the ansatz \eqref{ansatz-grav} together with a separable function $\Psi^V$
\be
\Psi^V=e^{i\omega t+im\phi}\Phi^V_{\omega\ell mn}(r)S^V_{\omega\ell mn}(\theta)\,,
\ee
the linearized Einstein equations lead to the following two ODEs
\be\ba\label{eq-graV}
{}&\frac{D^M_\theta}{\sin2\theta\cos^2\theta}\frac{d}{d\theta}\left[\frac{\sin2\theta\cos^2\theta}{D^M_\theta}\frac{d}{d\theta}S^V_{\omega\ell mn}\right]+\left[\frac{2\Lambda^V}{D_\theta^M}-2\Lambda^V-\frac{(m+a\omega\sin^2\theta)^2}{\sin^2\theta}\right] S^V_{\omega\ell mn}=0\,,\\
{}&\frac{D^M_r}{r^3}\frac{d}{dr}\left[\frac{r\Delta}{D^M_r}\frac{d}{dr}\Phi^V_{\omega\ell mn}\right]+\left[-\frac{2\Lambda^V}{D_r^M}+2\Lambda^V+\frac{R^2W^2}{r^2\Delta}\right]\Phi_{\omega\ell mn}=0\,,
\ea\ee
where $D_r^M,D_\theta^M$ are given by \eqref{choice} and 
\be
\ba
\Lambda^V=\mu\omega-\frac{am+a^2\omega}{\mu}\,,\quad 
\ea
\ee

\section{Static solutions and responses of master fields}\label{solution}
In this section, we solve the angular and radial equations in the static limit for perturbations discussed before.  
Then we study the asymptotic expansion of the master fields to find the static Love numbers and identify their vanishing conditions. 
Since $\omega$ is set to zero, we will omit it in the subscript of the functions involved for simplicity.
\subsection{Scalar perturbations}
We begin by a brief review of solving Klein-Gordon equation for scalar field as discussed in \cite{Charalambous:2023jgq}.
We first solve the angular equation. When $\omega=0$, the scalar angular equation becomes
\be\label{eq-ang}
\frac{1}{\sin2\theta}\frac{d}{d\theta}(\sin2\theta S'_{\ell mn})-\left(\frac{m^2}{\sin^2\theta}+\frac{n^2}{\cos^2\theta}\right)S_{\ell mn}=-\ell(\ell+2)S_{\ell mn}\,,
\ee
where we have parametrized $C_0$ as
\be
C_0=\ell(\ell+2)\,.
\ee
The solution to  \eqref{eq-ang} which is regular at $\theta=\pi/2$ is given by
\be\label{Ss}
S_{\ell mn}=\sin^{|m|}\theta\cos^{|n|}\theta{}_2F_1\left(\frac{-\ell+|m|+|n|}{2},\frac{\ell+|m|+|n|}{2}+1,1+|n|,\cos^2\theta\right)\,.
\ee
For $S$ to be regular at $\theta=0$ as well, $\ell$ has to be quantized such that
\be\label{quant:ell}
\frac{\ell-|m|-|n|}{2}\in\mathbb N\,.
\ee
Actually, the hypergeometric function in \eqref{Ss} reduces to a polynomial of $\cos\theta$ when \eqref{quant:ell} is satisfied. The quantization condition \eqref{quant:ell} can be achieved by requiring $\ell\in\mathbb N, |m|\leq\ell$ and $n$ to take integer values within 
\be
n=-(\ell-|m|)\,,\quad-(\ell-|m|)+2\,,\cdots,\ell-|m|-2\,,\quad\ell-|m|\,.
\ee
Now  we consider the radial equation and introduce the coordinate $z$ as\footnote{The definition of $z$ relates to $x$ used in \cite{Charalambous:2023jgq} by an $SL(2)$ transformation $z=\frac{x}{1+x}$ which does not affect the result.}
\be\label{z:rho}
z=\frac{\rho-\rho_+}{\rho-\rho_-}\,,\quad\rho=r^2\,,
\ee
where $\rho_\pm=r_\pm^2$ are related to the mass $M$ of the black hole via 
\be
M=\frac{(a^2+\rho_+)(b^2+\rho_+)}{\rho_+}\,,\quad \rho_+\geq\frac{a^2b^2}{\rho_+}\equiv \rho_-\,.
\ee
From \eqref{z:rho}, it is straightforward to see that $z=0$ is the outer horizon and $z=1$ is the radial infinity. In $z$ coordinate, the radial equation for the scalar field becomes
\be\label{eq:scalar-z}
\Phi''_{\ell mn}+\frac{1}{z}\Phi'_{\ell mn}+\left(\frac{A^2}{z^2}-\frac{C_0}{4(z-1)^2}+\frac{B}{z(1-z)}\right)\Phi_{\ell mn}=0\,,
\ee
where
\be\ba\label{parameter-scalar}
{}&A=\frac{A_++A_-}{2}\,,\quad B=A_+A_--\frac{C_0}{4}\,,
\quad A_\pm=\frac{\sqrt{\rho_+}M}{2(\rho_+-\rho_-)}(m\pm n)(\Omega_\phi\pm\Omega_\psi)\,.
\ea
\ee
Under the following field redefinition\footnote{We can also let $\Phi=e^{-iA}(1-z)^{\hat\ell+1}f$, so there are two branches of solutions which are labeled by $\sigma=\pm$ as has been done in \cite{Charalambous:2023jgq}. The vanishing condition for the Love number is independent of the sign and we will only focus on one branch for simplicity. }
\be\label{Phis}
\Phi_{\ell mn}=z^{iA}(1-z)^{\hat\ell+1} f(z)
\ee
with $\hat\ell=\frac{\ell}{2}$,
it can be shown that \eqref{eq:scalar-z} takes the canonical form of a hypergeometric equation for $f$
\be\label{eq:hyper}
f''+\frac{(a+b+1)z-c}{z(z-1)}f'+\frac{ab}{z(z-1)}f=0\,,
\ee
with
\be\label{coe-scalar}
a=\hat\ell+1+iA_+\,,\quad b=\hat\ell+1+iA_-\,,\quad c=1+2iA\,.
\ee
The solution which is regular at the future event horizon,  i.e. the ingoing branch, is simply given by
\be
f={}_2F_1(a,b,c,z)  \,,
\ee
where we have set the normalization factor to be 1.
The  scalar Love number $k_{\ell mn}^{S}$ for generic value of $\hat\ell$ is defined through the asymptotic behavior of $\Phi_{\ell mn}$ at infinity which is
\be\label{def:ks}
\Phi_{\ell mn}\sim \rho^{\hat\ell}\left(1+k^{S}_{\ell mn}\left(\frac{M}{\rho}\right)^{2\hat\ell+1}\right)\,.
\ee
Using $1-z=\frac{\rho_+-\rho_-}{\rho-\rho_-}\sim\frac{\rho_+-\rho_-}{\rho}$ and the expansion of the hypergeometric function
\be
{}_2F_1(a,b,c,z)\sim\frac{\Gamma(c)\Gamma(c-a-b)}{\Gamma(c-a)\Gamma(c-b)}-(1-z)^{c-a-b}\frac{\Gamma(1+a+b-c)\Gamma(c)}{(c-a-b)\Gamma(a)\Gamma(b)}\,,
\ee
it is easy to find that the scalar Love number is given by
\be\label{ks}
k_{\ell mn}^{S}=\left(\frac{\rho_+-\rho_-}{M}\right)^{2\hat\ell+1}\frac{\Gamma(-2\hat\ell-1)\Gamma(\hat\ell+1+iA_+)\Gamma(\hat\ell+1+iA_-)}{\Gamma(2\hat\ell+1)\Gamma(-\hat\ell+iA_+)\Gamma(-\hat\ell+iA_-)}\,.
\ee
However, since $\ell$ is quantized such that $2\hat\ell\in\mathbb Z_+$ according to \eqref{quant:ell}, the expression \eqref{ks} is actually divergent due to the factor $\Gamma(-2\hat\ell-1)$.  Actually, we should restrict $\hat\ell$ to be half integer before sending $z\to1$. The correct expansion of $\Phi_{\ell mn}$ around $z=1$ can be obtained using the following formula which is valid for $c\in\mathbb N$,
\be\ba\label{expand1}
{}&(-1)^{c-1}\frac{c!\Gamma(a-c)\Gamma(b-c)}{\Gamma(a+b-c)}{}_2F_1(a,b,a+b-c,z)
\\&={}_2F_1(a,b,c+1,1-z)\log(1-z)
-\sum_{n=1}^c\frac{c!(n-1)!}{(c-n)!(1-a)_n(1-b)_n}(z-1)^{-n}
\\&+\sum_{n=0}^\infty\frac{(a)_n(b)_n}{(c+1)_nn!}(\psi(a+n)+\psi(b+n)-\psi(1+n)-\psi(c+1+n))(1-z)^n\,,
\ea
\ee
where $(k)_n\equiv \frac{\Gamma(k+n)}{\Gamma(k)},\psi(z)\equiv\Gamma'(z)/\Gamma(z)$. For the solution we are interested in, $c=2\hat\ell+1$ is a positive integer. The logarithmic term is regarded as the running of the Love number and the inverse of coefficient in front of $(M/\rho)^{2\hat\ell+1}$ is identified as the $\beta$-function  associated with the RG flow of the Love number. As a result, instead of \eqref{def:ks}, we actually have
\be\label{Phi:beta}
\Phi_{\ell mn}\sim \rho^{\hat\ell}\left(1+\beta_{\ell mn}^{S}\left(\frac{M}{\rho}\right)^{2\hat\ell+1}\log\frac{\rho}{M}\right)\,,\quad\rho\to\infty
\ee
with $\beta_{\ell mn}^{S}$ given by
\be\label{rks}
\beta_{\ell mn}^{S}=\left(\frac{\rho_--\rho_+}{M}\right)^{2\hat\ell+1}\frac{\Gamma(\hat\ell+1+iA_+)\Gamma(\hat\ell+1+iA_-)}{(2\hat\ell)!(2\hat\ell+1)!\Gamma(-\hat\ell+iA_+)\Gamma(-\hat\ell+iA_-)}\,.
\ee
Be aware that $\beta^S_{\ell mn}$ depends not only on $\hat\ell$, but also on $m,n$ through $A_\pm$. It is also noted that although the result \eqref{ks} is ill defined for $2\hat\ell\in\mathbb Z_+$, it provides a quick way to compute the meaningful running Love number \eqref{rks} without using the expansion \eqref{expand1}. We can first expand the solution around $z=1$ with the  replacement $2\hat\ell\to2\hat\ell-\epsilon$. Using $\Gamma(-n+\epsilon)=\frac{(-1)^n}{n!\epsilon}+O(\epsilon^0)$ for $n\in\mathbb Z_+$,  the running Love number \eqref{rks} is equal to the residue of \eqref{ks} at $\epsilon=0$. 

The running Love number \eqref{rks} vanishes only if the denominator is divergent, which occurs when $\hat\ell\in\mathbb N$ and $A_+A_-=0$. The latter is satisfied when $|a|=|b|$ or $|m|=|n|$.

\subsection{Electromagnetic perturbations}\label{sec:EM-sol}

Having reviewed the computation of static Love number for scalar field, we are ready to study the electromagnetic master equations with $\omega=0$. We assume $m\geq0,n\geq0$ without loss of generality below for convenience. 
As we will see in the following, the two ODEs for electric polarization are identical to the scalar equations. However, for magnetic polarization, both the angular and radial differential equations become Heun equations. Surprisingly, these Heun equations are  special and can be solved in terms of hypergeometric functions. We will follow a similar route to the scalar case.  We begin with solving the angular equation. Imposing the regularity condition gives a quantization condition for the separation parameter $\mu$. Using such quantization condition, we can compute the logarithm response and determine the vanishing condition.

\subsubsection{Electric polarization}
As has been argued in \cite{Lunin:2017drx}, the static limit should be  taken such that 
$
\lambda_e\equiv\frac{\omega}{\mu}
$
is kept fixed. As a result, we have
\be
\Lambda=\lambda_e\,,\quad \mu=C=0\,,
\ee
and \eqref{eq} becomes an eigenvalue problem for $\lambda_e$. It is then straightforward to see that wave equations for electric polarization
are the same  as those for scalar perturbation under the identification
\be
\lambda_e=\frac{C_0}{2}=\frac{\ell(\ell+2)}{2}\,.
\ee
where $\ell$ is quantized to satisfy \eqref{quant:ell}.
Since $\ell\in\mathbb N$, the asymptotic expansion of the radial function has the same form as \eqref{Phi:beta} with the introduction of the running of static response $\beta_{\ell mn}^{\text{el}}$
whose value is the same as \eqref{rks}. Note that $\beta_{\ell mn}^{\text{el}}$ is related to, but  not identical to the Love number of the electric field in the context of EFT. To determine the latter, we need to reconstruct the electric field $A_t$ or the gauge invariant field strength $F_{tr}$. The static electric Love number is proportional to the response coefficient of $A_t$. This will be discussed in the next section.

\subsubsection{Magnetic polarization}
The discussion on magnetic polarization is more involved. We start by solving the angular equation. 
Parametrizing the separation constant $\mu$ as
\be\label{mu-l}
\mu=\frac{bm+an}{\ell}\,,
\ee
and letting
\be
S_{\ell mn}^{M}=x^{n/2}(1-x)^{m/2}\mathcal{S}_{\ell mn}^{M}(x)\,,
\ee
where $x=\cos^2\theta$,
the angular equation turns into a Heun equation \eqref{eq:heun} for $\mathcal{S}_{\ell mn}^{M}$ when $|a|\neq|b|$ with parameters being
\be\ba\label{parameters-mag}
{}&\gamma=1+n,\quad\delta=1+m,\quad\epsilon=-1,\quad \tilde q=\frac{(\mu-b)(\ell+m-n)}{2n(\mu-a)}\,,\\
{}&a_0=\frac{\mu^2-b^2}{a^2-b^2},\quad \alpha=-\frac{\ell-m-n}{2},\quad\beta=\frac{\ell+m+n}{2}\,.
\ea\ee
Interestingly, these coefficients satisfy \eqref{eq:q} and the solution which is regular at $x=0$ is given by
\be\label{Sm}
S_{\ell mn}^{M}(\theta)=x^{n/2}(1-x)^{m/2}\left[{}_2F_1(\alpha,\beta,n,x)+\tilde q ~{}_2F_1(\alpha,\beta,n+1,x)\right]\,.
\ee
Requiring the hypergeometric function to be regular at $x=1$ implies $\ell$ needs to satisfy $\frac{\ell-m-n}{2}\in\mathbb N$.
However, it can be checked that when $\ell=m+n$, the solution \eqref{Sm} is zero. Actually, by solving the angular equation directly with $\ell=m+n$, we find that there is no nontrivial solution regular at both $\theta=0$ and $\theta=\frac{\pi}{2}$. This suggests that the correct quantization condition for $\ell$ is
\be\label{quantcond}
{\frac{\ell-m-n}{2}\in\mathbb N_+}\,.
\ee
When $|a|=|b|$, the singular point $a_0\to\infty$ and the angular equation reduces to a hypergeometric equation.
It can be shown that solution for $|a|=|b|$ can be simply obtained from \eqref{Sm} by taking the limit, so we do not need to consider it separately. 
Solutions \eqref{Sm} associated with different eigenvalues $\ell$ are not orthogonal to each other for $|a|\neq |b|$. Actually, it can be shown that the eigenfunction $S^{M}_{\ell mn}$ can be written as a linear combination of $S_{\ell  mn}$ and $S_{(\ell +2) mn}$. As a result, the angular part of the master field for magnetic polarization can be expanded into orthogonal modified spherical harmonic basis. Details can be seen in appendix B.

Now we consider the radial equation. 
Using the coordinate transformation \eqref{z:rho} together with the following field redefinition
\be
\Phi_{\ell  mn}^{M}=z^{iA}(1-z)^{\frac{\ell }{2}}f^M_{\ell mn}\,,
\ee
the radial equation becomes Heun equation \eqref{eq:heun} for $f_m$ with coefficients being
\be
\ba\label{rad-para}
{}&\gamma=1+2iA\,,\quad \delta=1+\ell\, ,\quad\epsilon=-1\,,\quad \tilde q=-\frac{1}{2}+\frac{i\sqrt{\rho_+}}{2\mu}\left(\frac{A_-(\rho_++\mu^2)}{A(\rho_+-ab)}-1\right)\,,\\
{}&a_0=\frac{\rho_+(\rho_++\mu^2)}{a^2b^2+\mu^2\rho_+}\,,\quad\alpha=\frac{\ell }{2}+iA_+\,,\quad\beta=\frac{\ell }{2}+iA_-\,.
\ea
\ee
Values of $A_\pm$ and $A$ are given in \eqref{parameter-scalar}.
Similar to the angular equation, parameters \eqref{rad-para} for the radial Heun equation also satisfy the constraint equation \eqref{eq:q}, which suggests that the singularity $z=a_0$ is removable. As a result, the solution to the radial equation which is regular at the future horizon is given by
\be\label{PhiM}
\Phi_{\ell  mn}^{M}(r)=z^{iA}(1-z)^{\frac{\ell }{2}}\left[{}_2F_1\left(\frac{\ell }{2}+iA_+,\frac{\ell }{2}+iA_-,2iA,z\right)+\tilde q^M{}_2F_1\left(\frac{\ell }{2}+iA_+,\frac{\ell }{2}+iA_-,1+2iA,z\right)\right]\,.
\ee
Expanding the solution around $z=1$ 
 and taking the fact that $\ell \in\mathbb N$ into consideration,  the $\beta$-function associated to $\Phi_{\ell  mn}^{M}$ is introduced as follows
\be\label{km}
\Phi_{\ell  mn}^{M}\sim\rho^{\frac{\ell }{2}}\left(1+\beta_{\ell  mn}^{M}\left(\frac{M}{\rho}\right)^{\ell }\log\frac{\rho}{M}\right)\,,\quad\rho\to\infty\,.
\ee
Using the formula \eqref{expand1}, we have
\be\ba\label{betaM}
\beta_{\ell  mn}^{M}&=\left(\frac{\rho_--\rho_+}{M}\right)^{\ell }\frac{(\rho_++ab)(\ell ^2+4A_-^2)\Gamma\left(\frac{\ell }{2}+iA_+\right)\Gamma\left(\frac{\ell }{2}+iA_-\right)}{4(\rho_+-ab)\Gamma(\ell )\Gamma(\ell +1)\Gamma\left(1-\frac{\ell }{2}+iA_+\right)\Gamma\left(1-\frac{\ell }{2}+iA_-\right)}
\\&=\frac{M}{\rho_+-\rho_-}\frac{4\ell (\ell +1)(\rho_++ab)}{(\rho_+-ab)(\ell ^2+4A_+^2)}\beta_{\ell  mn}^{S}\,.
\ea
\ee
Therefore, it is easy to see that the vanishing condition for the running static response of magnetic polarization is similar to that of the scalar field, which is $\ell \in\mathbb N_+$ and $(a^2-b^2)(m^2-n^2)=0$.
\subsection{Gravitational perturbations}
We first solve the angular equation in \eqref{eq-graV}. In this case, we parametrize the separation constant $\mu$ as
\be\label{mu-l-grav}
\mu=\frac{2am}{\ell^2-1}\,.
\ee
Letting 
\be
S^V_{\ell mn}=(1-x)^{m/2}\mathcal{S}^V_{\ell mn}(x)\,,\quad x=\cos^2\theta\,,
\ee
the angular equation takes the form of a Heun equation \eqref{eq:heun} for $\mathcal{S}^V_{\ell mn}$ with relevant parameters being
\be
\ba\label{parameters-grav}
{}&\gamma=2\,,\quad \delta=1+m\,,\quad\epsilon=-1\,,\quad\tilde q=\frac{\mu m}{2(\mu-a)}\,, \\
{}&a_0=\frac{\mu^2}{a^2}\,,\quad \alpha=-\frac{\ell-m-1}{2}\,,\quad\beta=\frac{\ell+m+1}{2}\,.
\ea
\ee
One interesting observation is that \eqref{parameters-grav} can be obtained from \eqref{parameters-mag} by setting $b=0,n=1$. Therefore, the solution can be simply obtained which is 
\be
S^V_{\ell mn}=(1-x)^{m/2}\left[{}_2F_1(\alpha,\beta,1,x)+\frac{\mu m}{2(\mu-a)}{}_2F_1(\alpha,\beta,2,x)\right]\,,
\ee
where $\ell$ satisfies the quantization condition
\be
\frac{\ell-m-1}{2}\in\mathbb N_+\,.
\ee
Now let us solve the radial equation. As previous discussion, we perform a proper rescaling of $\Phi^V_{\ell mn}$ as following
\be
\Phi^V_{\ell mn}=z^{iA}(1-z)^{\frac{\ell+1}{2}}f^V_{\ell mn}\,,\quad z=\frac{\rho-\rho_+}{\rho}\,,
\ee
where $A=\frac{am}{2\sqrt{\rho_+}}$ is defined by \eqref{parameter-scalar} with $b=0$. Then
the radial equation in \eqref{eq-graV} becomes the Heun equation with parameters given by
\be\ba
{}&\gamma=1+2iA\,,\quad \delta=1+\ell\,,\quad\epsilon=-1\,,\quad\tilde q=-\frac{1}{2}+\frac{2iA}{(\ell+1)^2}-\frac{i(\ell-1)}{4A}\,,\\
{}&a_0=1+\frac{\rho_+}{\mu^2}\,,\quad\alpha=\frac{\ell+1}{2}+iA\,,\quad\beta=\frac{\ell-1}{2}+iA\,,
\ea
\ee
They satisfy the relation \eqref{eq:q} and therefore the solution which is regular at the future horizon is given by
\be
\Phi_{\ell mn}^V=z^{iA}(1-z)^{\frac{\ell+1}{2}}\left[{}_2F_1\left(\frac{\ell+1+2iA}{2},\frac{\ell-1+2iA}{2},2iA,z\right)+\tilde q{}_2F_1\left(\frac{\ell+1+2iA}{2},\frac{\ell-1+2iA}{2},1+2iA,z\right)\right]\,.
\ee
Using \eqref{expand1} and expanding the solution around $z=1$ gives the large $\rho$ behavior of $\Phi_{\ell mn}^V$
\be
\Phi_{\ell mn}^V\sim \rho^{\frac{\ell-1}{2}}\left(1+\beta^V_{\ell mn}\left(\frac{M}{\rho}\right)^\ell\log\frac{\rho}{M}\right)\,,\quad\rho\to\infty\,,
\ee
with the beta function given by
\be
\beta^V_{\ell mn}=\left(-\frac{M}{\rho_+}\right)^\ell\frac{(\ell-1)^2[(\ell+1)^2+4A^2]\Gamma\left(\frac{\ell+1+2iA}{2}\right)\Gamma\left(\frac{\ell-1+2iA}{2}\right)}{4(\ell+1)^2\Gamma\left(\ell\right)\Gamma\left(\ell+1\right)\Gamma\left(\frac{ 3-\ell+2iA}{2}\right)\Gamma\left(\frac{1-\ell+2iA}{2}\right)}\,.
\ee
The vanishing condition for the running Love number $\beta^V_{\ell mn}$ is then $A=0\,,\frac{\ell-1}{2}\in\mathbb N_+$, with the former condition implying $a=0$ or $m=0$ and the latter implying that $\ell$ is odd.

\section{ Love tensor of Myers-Perry black holes}
\subsection{Effective field theory and tidal Love numbers}
In the effective field theory description of the interaction between  a black hole and an external field at large distances, the black hole is modeled as a point particle \cite{Kol:2011vg}. 
When considering the quadratic coupling between the external field and the worldline of the particle, tidal Love number is defined as the coupling constant \cite{Porto:2016pyg,Nicolis:2017eqo,Wong:2019yoc}. 
Given a source of the external field, we can solve for its linear response due to the quadratic coupling, and the Love number can be read off from the asymptotic behavior of the source and response. The static response coefficients we computed in the last section are closely related to the tidal Love numbers in the EFT framework. Taking the scalar field as an example, the first term in \eqref{def:ks} is the source and the second term is the response. Modulo a pure numerical factor, $k_s$ is just the tidal Love number in the context of EFT. For the electromagnetic field,  the worldline coupling should be gauge invariant. Therefore, the Love number is encoded in the field strength $F_{\mu\nu}$ instead of $A_\mu$. This leads to two kinds of couplings which are electric and magnetic characterized by $F_{ti}$ and $F_{ij}$ respectively.
In the static limit,
the electric Love number can be read off from the asymptotic expansion of $A_t$ directly. Above discussions are valid for both non-rotating and rotating black holes. When the black hole is rotating, it is described as a spinning particle whose motion is governed by the MPD equation \cite{Mathisson:1937zz,Papapetrou:1951pa,Dixon:1970zza}. Furthermore, the interaction between the spinning particle and the electromagnetic field implies that modes with different $\ell$ are coupled with each other and $A_t$ cannot be written in the form of \eqref{def:ks}. This leads to a generalization of the tidal Love number to the tidal Love tensor which is generally defined through
\be\label{def:TLT}
A_t(r,\Omega)\sim \sum_{\ell,m,n}Y_{\ell mn}\left(\frac{\rho}{M}\right)^{\ell/2}\left(\mathcal{E}_{\ell mn}+\sum_{\ell',m',n'}\mathbf k_{\ell mn,\ell'm'n'}\mathcal{E}_{\ell' m'n'}\log\frac{\rho}{M}\left(\frac{M}{\rho}\right)^{\ell+1}\right)\,,
\ee
where the logarithm term arises due to the fact that $\ell$ is an integer. 
Similarly, the static magnetic response can be obtained by expanding the spatial components  $A_i$ and projecting it onto the vector harmonics on $S^3$. For the gravitational waves, the gauge invariant quantity which couples to the worldline is the Weyl tensor. The coupling can be classified into three types, leading to three kinds of responses: gravito-electric response, gravito-magnetic response, and tensor response. For a detailed review of various kinds of gauge invariant couplings and the definition of Love numbers in EFT, see \cite{Hui:2020xxx}.

In this paper, we focus on the electromagnetic excitations and compute the electric tidal Love tensor defined in \eqref{def:TLT}. The other kinds of couplings and their responses are left for future investigation.
Since there is no mixing between different $m$ and $n$, we have
\be
\mathbf k_{\ell mn,\ell'm'n'}=\mathbf k_{\ell mn,\ell'}\delta_{mm'}\delta_{nn'}\,.
\ee
Using the separation ansatz and the static solutions we find, the  gauge field $A_\mu$ for both electric and magnetic polarizations can be reconstructed, which allows us to compute the tidal Love tensor.  The coefficients $\mathcal{E}_{\ell mn}$ and the running Love tensor $\mathbf k_{\ell mn,\ell'}$ can be obtained by examining the asymptotic behavior of the projection of $A_t$ onto $Y_{\ell mn}$.

\subsection{Electric polarization}
The static solution to the master field of electric polarization is given by
\be\label{psiE}
\Psi^{E}(r,\theta,\phi,\psi)=\Phi^E_{\ell mn}(r)Y_{\ell mn}(\theta,\phi,\psi)\,,
\ee
with $Y_{\ell mn}$ given by  \eqref{sphebasis} and $\Phi^E_{\ell mn}$ given by
\be\ba\label{sol:elec}
\Phi^E_{\ell mn}(r)&=z^{iA}(1-z)^{\frac{\ell}{2}+1}{}_2F_1\left(\frac{\ell}{2}+1+iA_+,\frac{\ell}{2}+1+iA_-,1+2iA,z\right)\,,\quad z=\frac{r^2-\rho_+}{r^2-\rho_-}\,,
\ea\ee
where $A,A_\pm$ are given by \eqref{parameter-scalar}. 
Throughout this section, we suppress the subscripts $m$ and $n$ which are kept fixed and use  $\ell$ to label different modes of the master field, e.g.
$
\Phi^E_\ell:=\Phi^E_{\ell mn}\,,Y_{\ell}:=Y_{\ell mn}\,.
$
Plugging $\Psi^E$ into the ansatz \eqref{ansatz-general} with the choice \eqref{ansatz-e} and setting $\mu=0$ in the static limit gives
\be\label{At:elec}
A_t=\frac{e^{im\phi+in\psi}}{\rho+x_1^2}\left[2(\rho-\rho_+)(\rho-\rho_-)\frac{d}{d\rho}\Phi_{\ell }\mathbb S_{\ell }+(a^2-b^2)\sin\theta\cos\theta \Phi_{\ell }\mathbb S'_{\ell }\right]\,.
\ee 
Using the expansion \eqref{Phi:beta}, the radial function can be formally written as
\be\label{large-rho}
\Phi_\ell=\frac{\mathcal{E}_\ell}{\ell}\left(\frac{\rho}{M}\right)^{\frac{\ell}{2}}\left[f_1\left(\frac{M}{\rho}\right)+\beta^{S}_{\ell}\left(\frac{M}{\rho}\right)^{\ell+1}\log\frac{\rho}{M}f_2\left(\frac{M}{\rho}\right)\right]\,,
\ee
where $f_1(x)$ and $f_2(x)$ are functions of $x$ which admit a Taylor expansion around $x=0$ as
\be
f_i(x)=1+\sum_{k=1}^\infty f_i^{(k)}x^k\,,\quad i=1,2\,.
\ee
The relative coefficient $\beta^{S}_\ell$ in \eqref{large-rho} is the same as the scalar running Love number computed in \eqref{rks} and the overall factor $\mathcal{E}_\ell$ is irrelevant for the computation of tidal Love tensor.
Plugging \eqref{large-rho} into \eqref{At:elec}, we find that $A_t$ can be expanded as
\be\label{At-far}
A_t=e^{im\phi+in\psi}\sum_{p\in\mathbb N}\mathcal{K}^{(1)}_p(\theta)\left(\frac{\rho}{M}\right)^{\frac{\ell}{2}-p}+\mathcal{K}^{(2)}_p(\theta)\log\frac{\rho}{M}\left(\frac{M}{\rho}\right)^{\frac{\ell}{2}+1+p}\,.
\ee
It is easy to write down the expressions for leading coefficients $\mathcal{K}_0^{(1)}$ and $\mathcal{K}_0^{(2)}$ which are
\be
\ba\label{K0}
\mathcal{K}_0^{(1)}(\theta)=\mathcal{E}_\ell\mathbb S_{\ell}\,,\quad \mathcal{K}_0^{(2)}=-\frac{(\ell+2)\mathcal{E}_\ell\beta_\ell^S\mathbb S_\ell}{\ell}\,.
\ea
\ee
To work out the subleading coefficients $\mathcal{K}_{p\geq1}^{(i)}$, we need to compute $f_i^{(k)}$ for all $k\leq p$. For later use, we also give the result for $\mathcal{K}_1^{(1)}$ which is
\be\label{K1}
\mathcal{K}_1^{(1)}=\frac{\mathcal{E}_\ell}{\ell} \left[\left((\ell-2)f_1^{(1)}-\frac{\rho_++\rho_-+x_1^2}{M}\ell\right)\mathbb S_\ell+\frac{a^2-b^2}{M}\sin\theta\cos\theta\mathbb S_\ell'\right].
\ee
The coefficients $\mathcal{K}_p^{(i)}$ can be  expanded in terms of $\mathbb S_\ell$ as
\be\label{exp-K}
\mathcal{K}_p^{(i)}=\sum_{j}\mathcal{C}_p^{(i)j}\mathbb S_j\,,\quad i=1,2\,,
\ee
with $j=m+n,m+n+2,\cdots$. 
Plugging \eqref{exp-K} into \eqref{At-far} gives
\be
A_t=\sum_{p\in\mathbb N}\sum_j\mathcal{C}_p^{(1)j}Y_j(\Omega)\left(\frac{\rho}{M}\right)^{\frac{\ell}{2}-p}+\mathcal{C}^{(2)j}_pY_j(\Omega)\log\frac{\rho}{M}\left(\frac{M}{\rho}\right)^{\frac{\ell}{2}+1+p}\,.
\ee
Comparing with \eqref{def:TLT}, we find that
\be\label{Ej}
\mathcal{E}_{j}=\left\{\begin{array}{ll}
  \mathcal{C}_{\frac{\ell-j}{2}}^{(1)j}   &,\quad \ell\geq j\geq m+n  \\
 0    & ,\quad j>\ell 
\end{array}\right.,
\ee
and the  Love tensor $\mathbf k^E_{j,j'}$ associated with the electric polarization satisfies the following equation
\be\label{TLT}
 {
\sum_{j'}\mathbf k^E_{j,j'}\mathcal{E}_{j'}=\sum_{m+n\leq j'\leq\ell}\mathbf k^E_{j,j'}\mathcal{C}^{(1)j'}_{\frac{\ell-j'}{2}}=\mathcal{C}^{(2)j}_{\frac{j-\ell}{2}}\,,\quad j\geq\ell\,.}
\ee
Equation  \eqref{Ej} suggests that given a specific solution $\Psi^E$ with quantum numbers $\ell,m,n$, the sources of the electric field  with modes $m+n\leq j\leq\ell$ are activated, which can be used to solve the tidal Love tensor $\mathbf k^E_{j,j'}$ with $j\geq\ell,j'\leq\ell$ using \eqref{TLT}.  Therefore, we provide an iterative method to solve the tidal Love tensor $\mathbf{k}^E_{jj'}$ for $j\geq j'$, which quantifies how higher modes respond to the excitations of lower modes. 

In the following, we will explicitly compute the   Love tensor $\mathbf k^E_{j,j'}$ for $j'=m+n$ and $j'=m+n+2$.
To compute $\mathbf k^E_{j,m+n}$ for $j\geq m+n$, we let $\Psi^E$ be \eqref{psiE} with $\ell=m+n$.  In this situation, only the source with mode $\ell=m+n$ is excited.
Solving \eqref{TLT} for $\mathbf k^E_{j,m+n}$ gives
\be\label{k:m+n}
\mathbf k^E_{j,m+n}=\frac{\mathcal{C}^{(2)j}_{\frac{j-m-n}{2}}}{\mathcal{C}^{(1)(m+n)}_{0}}\,,\quad j\geq m+n\,.
\ee
The denominator  can be simply obtained by combining \eqref{K0} and \eqref{exp-K} with the result being
\be\label{C0:ell}
\mathcal{C}_0^{(1)\ell}=\mathcal{E}_{\ell}\,.
\ee
Note that above result holds for general $\ell$.
To compute $\mathcal{C}^{(2)j}_{\frac{j-m-n}{2}}$,
we only need to know the projection of $\mathcal{K}_p^{(2)}$ onto $\mathbb S_{m+n+2p}$ with $p=\frac{j-m-n}{2}$. This can be obtained by noting that $\frac{\mathbb S_{m+n+2p}}{\mathbb S_{m+n}}$ is a polynomial of degree $2p$ in $\cos\theta$. In general, we have
\be\label{Slp}
\frac{\mathbb S_{m+n+2p_1}}{\mathbb S_{m+n+2p_2}}=W(p_1,p_2)x^{p_1-p_2}\left(1+O(x^{-1})\right)\,,\quad x=\cos^2\theta\,,
\ee
where the coefficient $W(p_1,p_2)$ is given by
\be
W(p_1,p_2)=\frac{(-1)^{p_1-p_2}N_{m+n+2p_1}(m+n+p_1+1)_{p_1}}{N_{m+n+2p_2}(m+n+p_2+1)_{p_2}(n+p_2+1)_{p_1-p_2}}\,,\quad p_1\geq p_2\,,
\ee  
with $N_\ell$ being the normalization factor $N_{\ell mn}$ given by \eqref{norm}.
On the other hand, it can be  checked that $\frac{\mathcal{K}_p^{(2)}}{\mathbb S_{m+n}}$ happens to be a polynomial  of degree $2p$ in $\cos\theta$. 
The factor $\cos^{2p}\theta$ can only come from 
expanding the overall factor $\frac{1}{\rho+x_1^2}$ in \eqref{At:elec} in large $\rho$ limit and we only need to keep the term
\be\label{fac1}
\frac{1}{\rho}\left(\frac{-x_1^2}{\rho}\right)^p.
\ee
Multiplying it with terms in the bracket of \eqref{At:elec} and pick up the leading logarithm term, we find
\be
\mathcal{K}_p^{(2)}(\theta)=-\frac{\mathcal{E}_\ell(\ell+2)\beta_\ell^S}{\ell}\left(\frac{b^2-a^2}{M}\right)^p\mathbb S_\ell(\theta)x^p\left(1+O(x^{-1})\right)\,,\quad x=\cos^2\theta\,.
\ee
Consequently, $\mathcal{C}_p^{(2)(\ell+2p)}$, the coefficient of the projection of $\mathcal{K}_p^{(2)}$ onto $\mathbb S_{\ell+2p}$, can be obtained from \eqref{Slp} by setting $p_1=\frac{\ell+2p-m-n}{2},p_2=\frac{\ell-m-n}{2}$, 
\be\label{C2l}
\mathcal{C}^{(2)(\ell+2p)}_p=-\frac{\mathcal{E}_\ell(\ell+2)}{\ell}\beta_\ell^{S}\left(\frac{b^2-a^2}{M}\right)^{p}W\left(\frac{\ell+2p-m-n}{2},\frac{\ell-m-n}{2}\right)^{-1}\,.
\ee
Substituting \eqref{C0:ell} and \eqref{C2l} into \eqref{k:m+n} with $p=\frac{j-m-n}{2},\ell=m+n$, we find
\be
 {\mathbf k^E_{j,m+n}=-\frac{m+n+2}{(m+n)W(\frac{j-m-n}{2},0)}\left(\frac{b^2-a^2}{M}\right)^{\frac{j-m-n}{2}}\beta_{m+n}^{S}}\,.
\ee
In order to get $\mathbf k^E_{jmn,m+n+2}$, we 
solve \eqref{TLT} with $\ell=m+n+2$ for $\mathbf k^E_{j,m+n+2}$ and find
\be
\mathbf k^E_{j,m+n+2}=\frac{\mathcal{C}^{(2)j}_{\frac{j-m-n-2}{2}}-\mathbf k^E_{j,m+n}\mathcal{C}^{(1)m+n}_{1}}{\mathcal{C}^{(1)m+n+2}_0}\,,\quad j\geq m+n+2\,.
\ee
The coefficients $C_0^{(2)m+n+2}$ and $C^{(2)j}_{\frac{j-m-n-2}{2}}$ 
can be obtained in a similar way as before and they are simply given by \eqref{C0:ell} and \eqref{C2l} with $\ell=m+n+2,p=\frac{j-m-n-2}{2}$. 
Therefore, we only need to compute $\mathcal{C}_1^{(1)m+n}$, which is the coefficient of $\mathcal{K}_1^{(1)}$  projecting onto $\mathbb S_{m+n}$.  Using \eqref{K1} with $\ell=m+n+2$, we find
\be
\mathcal{C}_1^{(1)m+n}=\frac{(m+n+4)\mathcal{E}_{m+n+2}(b^2-a^2)}{(m+n+2)W(1,0)M}\,.
\ee
Putting everything together, the  Love tensor $\mathbf k_{jmn,m+n+2}$ is evaluated to be
\be
\ba
\mathbf k^E_{j,m+n+2}=&-\frac{m+n+4}{m+n+2}\left(\frac{b^2-a^2}{M}\right)^{\frac{j-m-n-2}{2}}
\\&\times\left(\frac{\beta_{m+n+2}^{S}}{W(\frac{j-m-n}{2},1)}-\left(\frac{b^2-a^2}{M}\right)^2\frac{(m+n+2)\beta^{S}_{m+n}}{(m+n)W(\frac{j-m-n}{2},0)W(1,0)}\right)\,.
\ea
\ee
\subsection{Magnetic polarization}
Thanks to the fact that the master equations for magnetic polarization can be solved in terms of hypergeometric functions, we can compute the tidal Love tensors analytically as well.
With indices $m$ and $n$ suppressed, we let the magnetic master field  be a single mode
\be\label{static-mag}
\Psi^{M}(r,\theta,\phi,\psi)=\Phi_\ell ^{M}(r)S^{M}_\ell (\theta)e^{im\phi+in\psi}\,,
\ee
where $\Phi^M_\ell$ and $S^M_\ell$ are given by \eqref{PhiM} and \eqref{Sm} respectively. Quantum numbers $\ell,m,n$ are subject to the quantization condition \eqref{quantcond}.
Substituting \eqref{static-mag} into the ansatz \eqref{ansatz-general} for magnetic polarization, we find
\be
\ba\label{At:mag}
A_t=\frac{e^{im\phi+in\psi}}{(\rho+\mu^2)(\rho+x_1^2)}\left[2(\rho-\rho_+)(\rho-\rho_-)\frac{d}{d\rho}\Phi^{M}_\ell  S^{M}_\ell +\Phi^{M}_\ell \mathcal{Y}_\ell (\rho,\theta)\right]\,,
\ea
\ee
where
\be
\mathcal{Y}_\ell =\frac{(\rho+\mu^2)(a^2-b^2)\sin\theta\cos\theta}{\mu^2-x_1^2}(S^{M}_\ell )'+\frac{(\rho+x_1^2)[ab(an+bm)-(am+bn)\mu^2]}{\mu(\mu^2-x_1^2)}S_\ell ^{M}\,.
\ee
Note that $\mu$ is given by \eqref{mu-l}.
In the large $\rho$ limit, the radial function $\Phi^{M}_\ell $ can be expanded as
\be\label{exp:Phimag}
\Phi_\ell ^{M}=\mathcal{E}_\ell ^{M}\left(\frac{\rho}{M}\right)^{\frac{\ell }{2}}\left[h_1(M/\rho)+\beta_{\ell }^{M}\left(\frac{M}{\rho}\right)^{\ell }\log\frac{\rho}{M}h_2(M/\rho)\right]\,,
\ee
where $\beta^M_\ell$ is given by \eqref{betaM} and  $h_i(x)$ are functions of $x$ with following Taylor expansions
\be
h_i(x)=1+\sum_{k=1}^\infty h_i^{(k)}x^k\,,\quad i=1,2\,.
\ee
Similar to the electric case, the overall factor $\mathcal{E}_\ell ^{M}$ is not important.
Plugging \eqref{exp:Phimag} into \eqref{At:mag}, we find that the gauge field can be formally written as
\be
\ba\label{At:mag:exp}
A_t=e^{im\phi+in\psi}\left(\sum_{p\in\mathbb N}\mathcal{M}_p^{(1)}(\theta)\left(\frac{\rho}{M}\right)^{\frac{\ell }{2}-1-p}+\mathcal{M}_p^{(2)}(\theta)\log\frac{\rho}{M}\left(\frac{M}{\rho}\right)^{\frac{\ell }{2}+1+p}\right)\,.
\ea
\ee
Expanding the angular function $\mathcal{M}_p^{(i)}$ in the modified spherical basis $\mathbb S_j$ as
\be\label{exp:ang:mag}
\mathcal{M}_p^{(i)}=\sum_j\mathcal{D}_p^{(i)j}\mathbb S_j\,,\quad i=1,2
\ee
with $j=m+n,m+n+2,\cdots$, then \eqref{At:mag:exp} becomes
\be
A_t=\sum_{p\in\mathbb N}\sum_j\mathcal{D}_p^{(1)j}Y_j\left(\frac{\rho}{M}\right)^{\frac{\ell }{2}-1-p}+\mathcal{D}_p^{(2)j}Y_j\log\frac{\rho}{M}\left(\frac{M}{\rho}\right)^{\frac{\ell }{2}+1+p}\,.
\ee
Comparing this with \eqref{def:TLT}, we get
\be\label{Ej-mag}
\mathcal{E}_j=\left\{
\begin{array}{ll}
 \mathcal{D}^{(1)j}_{\frac{\ell -j-2}{2}}    &  ,\quad m+n\leq j\leq \ell -2\\
  0   & ,\quad j>\ell -2
\end{array}
\right.\,,
\ee
and the Love tensor $\mathbf k^M_{jj'}$ satisfies the following equation
\be\label{TLT:magnet}
\sum_{j'}\mathbf k^M_{j,j'}\mathcal{E}_{j'}=\sum_{m+n\leq j'\leq\ell -2}\mathbf k^M_{j,j'}\mathcal{D}^{(1)j'}_{\frac{\ell -j'-2}{2}}=\mathcal{D}^{(2)j}_{\frac{j-\ell }{2}}\,,\quad j\geq\ell \,.
\ee
\eqref{Ej-mag} suggests that if we use \eqref{static-mag} as an input for the magnetic polarization, then sources of the electric coupling with modes $m+n\leq j\leq\ell-2$ are excited, and \eqref{TLT:magnet} implies that modes with $j\geq\ell$ will respond to such excitation. 
In particular, $\mathbf k^M_{jj'}$ is a strictly lower triangular matrix.

 We now present the concrete result of $\mathbf{k}^{M}_{jj'}$ for $j'=m+n$.
One might be concerned that it is difficult to expand the function $\mathcal{Y}_\ell $ in terms of $S_{\ell}$ due to the factor $\mu^2-x_1^2$ in the denominator. However, it can be directly checked that $\mathcal{Y}_\ell $ is a superposition of the basis $\mathbb S_\ell $ and $\mathbb S_{\ell -2}$ for modified spherical harmonics 
\be\ba\label{Ylam}
\mathcal{Y}_\ell =&y_1\frac{(a+b)(\ell ^2-(m+n)^2)}{N_\ell }S_\ell +y_2\frac{(a-b)((\ell -n)^2-m^2)}{N_{\ell -2}}S_{\ell -2}\,,
\ea\ee
where
\be
y_i=\frac{\ell }{4n(an+bm-a\ell )}\left(\rho+\mu^2+(-1)^i\frac{mn(a^2+b^2)+ab(m^2+n^2-\ell ^2)}{\ell ^2}\right)\,.
\ee
This makes the expansion \eqref{exp:ang:mag} and computation of tidal Love tensors tractable. To compute $\mathbf k^M_{j,m+n}$, we use $\Psi^{M}_{\ell }$ with $\ell =m+n+2$ as an input. Solving \eqref{TLT:magnet} gives
\be\label{k:mag:m+n}
\mathbf k_{j,m+n}^{M}=\frac{\mathcal D^{(2)j}_{\frac{j-m-n-2}{2}}}{\mathcal{D}^{(1)m+n}_{0}}\,.
\ee
Substituting \eqref{exp:Phimag} into \eqref{At:mag} and comparing with \eqref{At:mag:exp}, we find
\be
\mathcal{M}_0^{(1)}(\theta)=\frac{\mathcal{E}_\ell ^{M}}{M}\left[\ell  S_\ell ^{M}+\lim_{\rho\to\infty}\frac{\mathcal{Y}_\ell }{\rho}\right].
\ee
Using \eqref{Ylam} and \eqref{Slam}, it is easy to find 
\be\label{D0}
\mathcal{D}_0^{(1)\ell -2}=\frac{\mathcal{E}_\ell ^{M}}{MN_{\ell -2}}\frac{\ell (a-b)[(\ell -n)^2-m^2]}{2n(an+bm-a\ell )}\,.
\ee
Similar to the electric polarization, terms  needed to determine $\mathcal{D}^{(2)j}_{\frac{j-\ell }{2}}$ come from expanding the factor $\frac{1}{\rho+x_1^2}$ and keeping \eqref{fac1}. Finally, we find
\be
\ba\label{Dp}
\mathcal{D}_p^{(1)\ell +2p}=\frac{\mathcal{E}_\ell ^{M}\beta_\ell ^{M}}{MN_\ell}\left(\frac{b^2-a^2}{M}\right)^p\frac{\ell (a+b)[\ell ^2-(m+n)^2]}{2  n(an+mb-a\ell )}W\left(\frac{\ell +2p-m-n}{2},\frac{\ell -m-n}{2}\right)^{-1}\,.
\ea
\ee
Therefore, plugging \eqref{D0} and \eqref{Dp} into \eqref{k:mag:m+n} with $\ell =m+n+2,p=\frac{j-m-n-2}{2}$ gives
\be
{\mathbf k_{j,m+n}^{M}=\left(\frac{b^2-a^2}{M}\right)^{\frac{j-m-n-2}{2}}\frac{(a+b)(m+n+1)N_{m+n}}{(a-b)(m+1)N_{m+n+2}}W\left(\frac{j-m-n}{2},1\right)^{-1}\beta_{m+n+2}^{M}}\,.
\ee
For generic $j,j'$, the Love tensor $\mathbf k_{j,j'}^{M}$ can be computed iteratively by solving \eqref{TLT:magnet}.

\section{Near Zone approximation}
In this section, we consider the near zone approximation of the radial equation for magnetic polarization. We start with a brief review of this approximation and the emergence of Love symmetry for massless scalar wave equation following \cite{Charalambous:2023jgq}.
\subsection{Massless scalar field}
The near-zone regime is defined as
\be\label{near}
\omega(r-r_+)\ll1\,,\quad \omega r_+\ll1\,.
\ee
In this regime, the wave equation for massless scalar admits an approximation by neglecting subleading terms in a proper way. Consequently, the radial equation can be identified with the Casimir equation of an SL$(2)$ algebra, indicating the emergence of SL$(2)$ Love symmetry.  
To be more precise, we  write \eqref{eq} as
\be
\ba
\mathbb P_{\text{full}}S^S_{\omega\ell mn}=0\,,\quad \mathbb O_{\text{full}}\Phi^S_{\omega\ell mn}=0\,.
\ea
\ee
For the scalar case, the operator $\mathbb O_{\text{full}}$ can be written into the following form
\be
\mathbb O_{\text{full}}=\frac{d}{d\rho}\Delta\frac{d}{d\rho}+V=\frac{d}{d\rho}\Delta\frac{d}{d\rho}+V_0+V_1,\quad \Delta=(\rho-\rho_+)(\rho-\rho_-)\,,
\ee
where  we have split the potential $V$ into two parts $V_0,V_1$ such that $V_0\gg V_1$ in the near zone regime \eqref{near}.
This splitting is not unique and one can modify $V_0$ with any term which is subleading. It is not difficult to show that in order for the approximation to be valid, the residue of $V_0$ at $\rho=\rho_+$ should agree with that of $V$, i.e.
\be\label{Res}
\text{Res}_{\rho\to\rho_+}V_0=\text{Res}_{\rho\to\rho_+}V\,.
\ee
At $\rho=\rho_-$, the residue of $V_0$ could differ from that of $V$ by terms of order $\omega$. One motivation for the near zone approximation is to have a non-static wave equation which can be solved analytically at the leading order. Then we can treat $V_1$ as a perturbation and solve the wave equation order by order.
For the scalar perturbation, the leading order ODE 
\be\label{leading}
\frac{d}{d\rho}\Delta\frac{d}{d\rho}\Phi^S_{\omega\ell mn}+V_0\Phi^S_{\omega\ell mn}=0
\ee
can be made a hypergeometric differential equation if $V_0$ is chosen to have the following form
\be
V_0=\frac{\mathcal{R}_+(m,n)}{\rho-\rho_+}+\frac{\mathcal{R}_-(m,n)}{\rho-\rho_-}+\mathcal{R}_0\,,
\ee
where $\mathcal{R}_+$ is fixed by the condition \eqref{Res} and given by
\be\label{res+}
f_1(m,n)=\frac{\rho_+M^2}{4(\rho_+-\rho_-)}(\omega+m\Omega_\phi+n\Omega_\psi)^2\,.
\ee
In \cite{Charalambous:2023jgq}, the authors chose
\be
\mathcal{R}_-(m,n)=-\mathcal{R}_+(n,m),\quad \mathcal{R}_0=-\frac{\ell(\ell+2)}{4}\,.
\ee
It is a valid approximation which can be proved by directly checking $V-V_0\ll V_0$. The solution to the leading order radial differential equation is given by
\be
\Phi^S_{\omega\ell mn}=z^{iA^{(\omega)}}(1-z)^{\frac{\ell}{2}+1}{}_2F_1\left(\frac{\ell}{2}+1+iA_+^{(\omega)},\frac{\ell}{2}+1+iA_-^{(\omega)},1+2iA^{(\omega)},z\right)\,,
\ee
where
\be
A_\pm^{(\omega)}=\frac{\beta_H}{4}[(1\pm 1)\omega+2(m\pm n)(\Omega_\phi\pm\Omega_\psi)]\,,\quad A^{(\omega)}=\frac{A_+^{(\omega)}+A_-^{(\omega)}}{2}\,.
\ee
Besides being solvable at the leading order, the near zone approximation also features the interesting property that the wave equation exhibits an SL$(2)$ symmetry.
More explicitly, if we define
\be
\ba
L_0&=-\beta_H(\p_t+\Omega_+\p_+)\,,
\\L_{\pm1}&=e^{\pm t/\beta_H}\left[\mp\sqrt{\Delta}\p_\rho+\p_\rho(\sqrt{\Delta})\beta_H(\p_t+\Omega_+\p_+)+\frac{\rho_+-\rho_-}{2\sqrt{\Delta}}\beta_H\Omega_-\p_-\right]\,,
\ea
\ee
where $\beta_H=\frac{\sqrt{\rho_+}M}{\rho_+-\rho_-},\Omega_\pm=\Omega_\psi\pm\Omega_\phi$ and $\p_\pm=(\p_\psi\pm\p_\phi)/2$\,, then they satisfy SL$(2)$ algebra
\be
[L_m,L_n]=(m-n)L_{m+n}\,,\quad m,n=0,\pm1\,.
\ee
The radial equation \eqref{leading} for $\Phi^S_{\omega\ell mn}$ is proved to be equivalent to the Casimir equation for $\Psi^S$ given by \eqref{Psi}, which is
\be
\left[L_0^2-\frac{1}{2}(L_1L_{-1}+L_{-1}L_1)\right]\Psi^S=\ell(\ell+1)\Psi^S\,.
\ee
The emergence of SL$(2)$ symmetry for the wave equation in the near zone regime is very similar to the story of Kerr/CFT correspondence, where  \cite{Guica:2008mu,Castro:2010fd}. 

\subsection{Magnetic polarization}
In the following, we study the near zone approximation of electromagnetic field wave equations. In particular, we focus only on the magnetic polarization.  In this case, the operator $\mathbb O_{\text{full}}$ is given by
\be
\mathbb O_{\text{full}}=D_\rho\frac{d}{d\rho}(\frac{\Delta}{D_\rho}\frac{d}{d\rho})+V\,,\quad D_\rho=1+\frac{\rho}{\mu^2}\,.
\ee
The residue of the potential $V$ at $\rho=\rho_+$ is the same as \eqref{res+},
\be
\text{Res}_{\rho\to\rho_+}V=\frac{\rho_+M^2}{4(\rho_+-\rho_-)}(\omega+m\Omega_\phi+n\Omega_\psi)^2\,.
\ee
Similar to the scalar case, we engineer the splitting of the potential such that the leading order near zone equation can be solved analytically. As we have studied in previous sections, we let the approximated equation to be a Heun equation with one fake singularity.
In order for the approximated ODE to be a Heun equation, we can choose the leading potential to take the following form
\be
V_0=\frac{\mathcal{R}_+(m,n)}{\rho-\rho_+}+\frac{\tilde{\mathcal{R}}_-(m,n)}{\rho-\rho_-}+\frac{\mathcal{R}_\mu}{\rho+\mu^2}+\mathcal{R}_0\,,
\ee
with $\mathcal{R}_+$ given by \eqref{res+}. The  other functions are required to make  $V-V_0=O(\omega)$. If we also require the approximated Heun equation to be solvable, then $\tilde{\mathcal{R}}_-,\mathcal{R}_\mu$ and $\mathcal{R}_0$ need to satisfy the following constraint equation
\be
\mathcal{R}_0=\frac{(\rho_-+\mu^2)\mathcal{R}_++(\rho_++\mu^2)\tilde{\mathcal R}_--\mathcal{R}^2_\mu}{(\rho_++\mu^2)(\rho_-+\mu^2)}\,.
\ee
There are still infinite choices left and we give one of them in the following which leads to  relatively simple solution. We let
\be\ba
\tilde{\mathcal{R}}_-&=-\frac{\rho_+M^2}{4(\rho_+-\rho_-)}(\omega+m\Omega_\phi+n\Omega_\psi)^2=-\mathcal{R}_-\,,\\
\mathcal{R}_\mu&=\frac{ab(an+bm)}{2\mu}-\frac{\mu(am+bn)}{2}-\frac{M\mu\omega }{2}\,,\\
\mathcal{R}_0&=-\frac{(an+bm)^2}{4\mu^2}=-\ell ^2_\omega\,,
\ea\ee
which can be easily checked to be a valid near zone approximation.
The approximated radial equation 
\be\label{leading-mag}
D_\rho\frac{d}{d\rho}\left(\frac{\Delta}{D_\rho}\frac{d}{d\rho}\right)\Phi^M_{\omega\ell mn}+V_0\Phi^M_{\omega\ell mn}=0
\ee
can be solved analytically with the solution being
\be\label{sol-omega}
\Phi^M_{\omega\ell mn}=z^{iA_\omega}(1-z)^{\ell _0}\left({}_2F_1(\ell _0+i\tilde A_+^\omega,\ell _0+i\tilde A^\omega_-,2iA_\omega,z)+\tilde q_\omega{}_2F_1(\ell _0+iA_+^\omega,\ell _0+iA^\omega_-,1+2iA_\omega,z)\right)\,
\ee
where
\be
\tilde A^\omega_\pm=A_\omega\pm\tilde A'_\omega\,,\quad\tilde A'_\omega=\frac{\sqrt{\rho_+}M}{2(\rho_+-\rho_-)}(n\Omega_\phi+m\Omega_\psi+\sqrt{\rho_-/\rho_+}\omega)\,,\quad
\tilde q_\omega=-\frac{1}{2}+\frac{i\sqrt{\rho_+}}{2\mu}\left(\frac{\tilde A_-^\omega(\rho_++\mu^2)}{A_\omega(\rho_+-ab)}-1\right)\,.
\ee
Therefore, we have obtained a non-static approximated solution to the electromagnetic wave in the near zone regime.
We conclude this section with a remark.
We have been dealing with Heun equations with one removable singularity. The solution is given by a sum of two hypergeometric functions. Motivated by the structure of the solution, 
it can be shown that there exists a field redefinition such that the Heun equation can be transformed into a hypergeometric equation.
When black hole wave equations reduce to hypergeometric form, it often indicates an underlying SL$(2)$ symmetry acting on the solution space, typically emerging in near-horizon or low-energy limits. Therefore,
it would be interesting to explore whether the near zone approximation for the magnetic polarization makes the leading order wave equation \eqref{leading-mag} enjoy any emergent symmetry, similar to the scalar case. We leave this for future investigation.
\section{Conclusions}
In this paper, we study wave equations for electromagnetic and gravitational perturbations in the 5-dimensional Myers-Perry black hole. For the electromagnetic perturbation, \cite{Lunin:2017drx} showed that under proper separation ansatz, Maxwell equation can be transformed into two decoupled ODEs for a scalar master field. Depending on the choice of the ansatz, the electromagnetic perturbation can be classified into two types: electric polarization and magnetic
polarization. For the gravitational perturbation, \cite{Lunin:2025yth} showed that when one of the rotating parameters of the black hole is set to 0, the linearized Einstein equations for the vector type of the perturbation are also separable. In the end, one arrives at decoupled angular and radial master ODEs for each kind of  perturbation. They define eigenvalue problems of the separation constant $\mu$.

We solve these wave equations in the static limit. The master equations of the electric polarization are the same as those of a massless scalar field. The angular equation is solved by the modified spherical harmonics and the radial equation is solved by the hypergeometric function. Master equations of both magnetic polarization and gravitational perturbation are Heun equations. The parameters of these Heun equations are so special that one of the regular singular points is removable. As a result, they can be solved analytically as the sum of two hypergeometric functions. By solving the angular equations, we find that the separation parameter $\mu$ is not arbitrary but related to the quantized orbital number $\ell$ via \eqref{mu-l} for the magnetic polarization
or \eqref{mu-l-grav}  for the gravitational perturbation. By solving the radial equations and expanding solution at far infinity, we can access the asymptotic behavior of the master fields which allows us to read off the running Love number $\beta_\ell^M$ and $\beta_\ell^V$. 

We then study the static tidal response defined within the framework of EFT using the solutions we have obtained. In particular, we focus on the electromagnetic perturbations and compute the corresponding responses defined through the electric coupling. In order to do so, we reconstruct the gauge field $A_t$ using the master field we have solved. By expanding $A_t$ in the large $\rho$ limit, we find a mixing between modes of sources and responses, leading to a tensor structure of the Love number. For each kind of polarization, we provide an iterative way of computing such Love tensors $\mathbf{k}^E_{j,j'}$ or $\mathbf{k}^M_{j,j'}$. We find that the Love tensors are lower triangular matrices, implying that higher modes will be activated in response to the excitation of lower modes.
We also explicitly present results of these Love tensors for the first few terms. Finally, we also discuss the near zone approximation of the radial wave equation for the magnetic polarization. Like the scalar case, there are infinite choices of the approximation in principle. We explicitly provide one choice such that the leading order wave equation can be solved analytically.

Besides exploring whether the leading order wave equation for the magnetic polarization in the near zone approximation has any hidden symmetry structure,
there are also some other questions that can be studied in the future. The separability of wave equations for electromagnetic fields in general dimensional Myers-Perry black holes has been shown in \cite{Lunin:2017drx}, and it would be interesting to see if the corresponding master equations can be solved analytically. There are also other kinds of excitations whose wave equations are separable, including massive gauge fields \cite{Frolov:2018pys} and higher form fields \cite{Lunin:2019pwz}. Moreover, one can also study the tidal responses of other kinds of couplings by projecting the gauge invariant quantities onto vector or tensor harmonics.

\subsection*{Acknowledgments}
We thank Bin Chen and Joan Sim\'on for helpful discussions.
\appendix
\section{Heun equation}
The Heun equation is a second order ODE with four regular singular points at $z=0,1,a_0,\infty$. It can be written in the following  form
\be\label{eq:heun}
f''+\left(\frac{\gamma}{z}+\frac{\delta}{z-1}+\frac{\epsilon}{z-a_0}\right)f'+\frac{\alpha\beta z-q}{z(z-1)(z-a_0)}f=0\,,
\ee
where the parameters satisfy the Fuchsian relation $1+\alpha+\beta=\gamma+\delta+\epsilon$. 
It is usually difficult to work out the closed form for the connection formula of the Heun function. However, there are special cases when one of the singularities is apparent and the Heun equation can be solved in terms of the hypergeometric functions. Let us consider the possibility when $z=a_0$ is an apparent singularity so that the solution has trivial holonomy around this point. This happens when $\epsilon=-N$ is a negative integer. Trivial holonomy around implies that we can perform a Taylor expansion of $f$ around $z=a_0$
\be\label{taylor}
f=\sum_{i=0} c_i(z-a_0)^i\,\quad c_0=1.
\ee
Plugging this expansion into \eqref{eq:heun} and solving the equation order by order, one can determine coefficients $c_i$. At order $(z-a_0)^{-N-1}$, 
the Heun equation leads to an algebraic equation $P_{N+1}[q]=0$ for $q$ where $P_{N+1}[q]$ is a specific polynomial of degree $N+1$ with coefficients being independent of $c_i$. 
When $q$ satisfies such constraint equation, the expansion \eqref{taylor} is consistent with the Heun equation and  $z=a_0$ is a removable singularity. Moreover, it is shown that
the solution can be written as a finite sum of hypergeometric functions \cite{Ishkhanyan:2014wma}. For the case we are interested in, we will present the result for $\epsilon=-1$. In this case, it can be shown that the constraint equation for $q$ is
\be
\ba\label{eq:q}
(q-a_0\alpha\beta+a_0(1-\delta))(q-a_0\alpha\beta+(a_0-1)(1-\gamma))=a_0(1-a_0)(1+\alpha-\gamma)(1+\beta-\gamma)\,.
\ea
\ee
When \eqref{eq:q} is satisfied, 
the solution to \eqref{eq:heun} is given by
\be\label{heun:sol}
f={}_2F_1(\alpha,\beta,\gamma-1,z)+\tilde q{}_2F_1(\alpha,\beta,\gamma,z)\,,
\ee
where 
\be\label{tildeq}
\tilde q=\frac{q-a_0\alpha\beta+a_0(1-\delta)}{(1-a_0)(\gamma-1)}\,.
\ee

\section{Modified Spherical Harmonics basis}
As has been noted in section \ref{sec:EM-sol},  the angular eigenfunction of the magnetic polarization with different eigenvalues $\ell$ are not orthogonal to each other. Actually, the solutions are sums of two modified spherical harmonics basis on $S^3$. In the Hopf coordinates $(\theta,\phi,\psi)$, the metric of $S^3$ is given by
\be
ds^2_{S^3}=d\theta^2+\sin^2\theta d\phi^2+\cos^2\theta d\psi^2\,.
\ee
A more familiar representation of $S^3$ uses the spherical coordinates $(\theta_1,\theta_2,\varphi)$ with the metric being
\be
ds^2_{S^3}=d\theta_1^2+\sin^2\theta_1(d\theta_2^2+\sin^2\theta_2d\varphi^2)\,.
\ee
The two coordinates are related via the following transformation
\be
\sin\theta=\sin\theta_1\sin\theta_2\,,\quad\tan\psi=\tan\theta_1\cos\theta_2\,,\quad \phi=\varphi\,.
\ee
We will be using the Hopf coordinates below and
the  modified scalar spherical harmonics basis $Y_{\ell mn}$ on $S^3$ satisfies the following differential equations
\be\label{eq:Ylmn}
\Delta_{S^3}Y_{\ell mn}=-\ell(\ell+2)Y_{\ell mn}\,,\quad \p_\phi Y_{\ell mn}=imY_{\ell mn}\,,\quad\p_\psi Y_{\ell mn}=inY_{\ell mn}\,.
\ee
Solving the eigenvalue equations \eqref{eq:Ylmn} directly gives 
\be\label{sphebasis}
Y_{\ell mn}(\theta,\phi,\psi)=e^{im\phi+in\psi}\mathbb S_{\ell mn}(\theta)\,,
\ee
where
\be
\mathbb S_{\ell mn}(\theta)=N_{\ell mn}\sin^m\theta\cos^n\theta{}_2F_1\left(-\frac{\ell-m-n}{2},\frac{\ell+m+n}{2}+1,n+1,\cos^2\theta\right)\,.
\ee
Regularity at $\theta=0$ implies $\frac{\ell-m-n}{2}\in\mathbb N$.
The normalization factor $N_{\ell mn}$ is chosen to be
\be\label{norm}
N_{\ell mn}=\frac{1}{2\pi n!}\sqrt{(2\ell+2)\frac{\left(\frac{\ell+m+n}{2}\right)!\left(\frac{\ell-m+n}{2}\right)!}{\left(\frac{\ell-m-n}{2}\right)!\left(\frac{\ell+m-n}{2}\right)!}}\,,
\ee
so that $Y_{\ell mn}$ satisfies the standard orthonormal condition on  $S^3$
\be
\int d\Omega_3Y^*_{\ell mn}Y_{\ell' m'n'}=\delta_{\ell\ell'}\delta_{mm'}\delta_{nn'}\,,
\ee
where  the integral is defined as
\be
\int d\Omega_3=\int_0^{\frac{\pi}{2}}d\theta\int^{2\pi}_0d\phi\int^{2\pi}_0d\psi\sin\theta\cos\theta\,.
\ee
Now we consider the solution \eqref{Sm} to angular equation for the magnetic polarization. Using Gauss's contiguous relations, $S^M_{\ell mn}$  can be written as
\be\label{Slam}
S_{\ell  mn}^{M}=\frac{(a+b)(m+n+\ell )(m+n-\ell )}{4N_{\ell mn}n(an+bm-a\ell )}\mathbb S_{\ell  mn}+\frac{(b-a)(m-n+\ell )(m+n-\ell )}{4N_{(\ell-2)mn}n(an+bm-a\ell )}\mathbb S_{(\ell -2)mn}\,.
\ee
Therefore, if we define $Y_{\ell  mn}^{M}$ as the angular function of the master field $\Psi^M$
\be
Y_{\ell  mn}^{M}(\theta,\phi,\psi)=e^{im\phi+in\psi}S_{\ell  mn}^{M}(\theta)\,,
\ee
 then this angular function can be written as the superposition of two modified spherical harmonics in the following way
 \be
 Y_{\ell  mn}^{M}=\frac{(a+b)(m+n+\ell )(m+n-\ell )}{4N_{\ell  mn}n(an+bm-a\ell )}Y_{\ell  mn}+\frac{(b-a)(m-n+\ell )(m+n-\ell )}{4N_{(\ell -2) mn}n(an+bm-a\ell )}Y_{(\ell -2)mn}\,.
 \ee

\bibliographystyle{JHEP}
\bibliography{LN}
\end{document}